\documentclass{amsart}
\usepackage{amsthm}
\usepackage{amssymb}
\usepackage{amsmath}
\usepackage{graphicx}

\theoremstyle{definition}

\theoremstyle{remark}

\expandafter\chardef\csname pre amssym.def at\endcsname=\the\catcode`\@
\catcode`\@=11
\def\undefine#1{\let#1\undefined}
\def\newsymbol#1#2#3#4#5{\let\next@\relax
 \ifnum#2=\@ne\let\next@\msafam@\else
 \ifnum#2=\tw@\let\next@\msbfam@\fi\fi
 \mathchardef#1="#3\next@#4#5}
\def\mathhexbox@#1#2#3{\relax
 \ifmmode\mathpalette{}{\m@th\mathchar"#1#2#3}%
 \else\leavevmode\hbox{$\m@th\mathchar"#1#2#3$}\fi}
\def\hexnumber@#1{\ifcase#1 0\or 1\or 2\or 3\or 4\or 5\or 6\or 7\or 8\or
 9\or A\or B\or C\or D\or E\or F\fi}

\font\teneufm=eufm10
\font\seveneufm=eufm7
\font\fiveeufm=eufm5
\newfam\eufmfam
\textfont\eufmfam=\teneufm
\scriptfont\eufmfam=\seveneufm
\scriptscriptfont\eufmfam=\fiveeufm

\catcode`\@=\csname pre amssym.def at\endcsname

\newcommand{\cc}{\c{c}}

\newcommand{\ii}{\'{\i}}

\def    \integers       {{\mathbb Z}}

\def    \p      {\partial}

\begin{document}




\noindent
\begin{center}
{\Large{{\bf The Spherical Landau Problem}}} 
\end{center}
\vspace{1.0cm}
\begin{center}
{{{\bf{C\'esar Castilho}}\footnote{castilho@dmat.ufpe.br}\\{\it Depto. Matem\'atica - Universidade
Federal de Pernambuco, 
Recife - PE, Brazil \\
The Abdus Salam International Centre for Theoretical Physics \\(Visiting Scientist)\\ 
\vspace{0.5cm}
{\bf{Andr\'e Penna-Firme}}\footnote{apennafi@ictp.trieste.it}\footnote{Corresponding author}\\
{\it  The Abdus Salam International Centre of 
Theoretical Physics - (Visiting Scientist)\\Strada Costiera 11 - 34014, Trieste -Italy}\\Conselho Nacional 
de Desenvolvimento Cient\ii fico e Tecnol\'ogico
CNPq-Brazil\\Faculdade de Educa\cc \~ao, Federal University of 
Rio de Janeiro, (FE/UFRJ)
}}}
\end{center}

\date{October, 2001}
\thanks{APF thanks CNPq-Brazil for finantial support}

\vspace{2.0cm}
{\centerline {\bf Abstract}}
\vspace{0.5cm}
 The magnetization for electrons on a 
two-dimensional sphere, under a spherically symmetrical normal magnetic 
field has been studied in the large field limit. This allows
us to use an Euclidean approximation for low energies electron states
getting an analytical solution for the problem and 
avoiding the difficulties of quantization on a curved manifold. At low
temperatures our results are exact and allow direct
comparisson with the planar Landau case. In this temperature limit
we compute the magnetization and show it exhibit an oscillatory {\it de Hass-Van
Alphen}
type of behaviour. 

\vspace{1.0cm}
\newpage
\section{Introduction}
Landau studied the problem of the magnetization of a system of free-electrons 
in the presence of a perpendicular magnetic field in Euclidean three-space(\cite{La}). It is
well known that, at low temperatures and in the limit of a strong magnetic field
$B$, many magnetic properties of the system  show an oscilatory behaviour 
as a function of $\frac{1}{B}$(\cite{ca}). Typical examples are the {\em de Haas-van Alphen
effect} and the {\em Shubnikov-de Haas effect}. The first one refers to the 
magnetic susceptibility while the other to the magnetoresistance. 
Those effects are described
 with the simple model of free electrons with an effective mass $\mu$. 
The electron spin
coupling with the external magnetic field and spin-orbit contributions are
approximately incorporated through an effective scalar giromagnetic $g$
factor. Boundary effects have been considered in several papers
(see references whithin(\cite{Fu})) and the effects of confining potentials were studied by Kubo
(\cite{ku})
at the Low temperature limit. More recently (\cite{Fu}) dealt with 
the problem of current distribution under harmonic confining potentials.

In this work our main goal is to study the effects of the curvature of a
two-dimensional  substract in the orbital magnetic properties in the limits 
of strong field and low
temperature. For the sake of simplicity we assume that the curvature is constant and
positive, that is, we consider the case of a normal magnetic field
to a 2-dimensional sphere.  The high field limit allows 
one to work in a semilocal
approximation. This is due to the fact that the classical motion of the
 electron in the mentioned limit is, for small values of energy, confined
to a 
small neighborhood of its initial position. The classical canonical 
quantization for the Euclidean
plane can then be invoked as a first order approximation for the problem. 
As it will be shown, the first order terms of our results coincides with 
the standards results for the planar Landau problem(\cite{ca}). 

The paper is organized as follows. In section (\ref{quant}) we built our model
and compute the energy eigenvalues. In section(\ref{mag}) we compute the free-energy.
In section (\ref{curv}) we study the magnetization of our system and show 
that a {\it de Hass-Van Alphen effect} is present. In section(\ref{conc}) we 
draw our final conclusions.

\section{The Quantization}
\label{quant}
The classical hamiltonian of a charged particle of charge $e$, with effective mass
$\mu$ on a sphere of radius $r$ under a constant normal
magnetic field is given by
\begin{equation}
\label{1h}
 H = \frac{(p_{\theta}- e \, A_{\theta})^2}{2\,\mu \,  r^2} +
\frac{(p_{\phi}- 
e\, A_{\phi})^2}{2 \, \mu \, r^2 \, \cos^2(\theta)}. 
\end{equation}
Here $(\phi, \theta )$ are standard spherical coordinates and 
the magnetic field is given by
$$ \vec B(\theta, \phi) = \nabla \times \vec A(\theta, \phi), $$
and $A_{\theta}, \, A_{\phi}$ are such that 
$  \vec B = b \, \hat r . $ We work in the gauge
$$ A_{\theta} = 0 \, \,  \mbox{and} \, \, A_{\phi}= b \,r \,\theta. $$ 
The main reason for this gauge choice is that the Hamiltonian becomes
$\phi$ independent and therefore separable. By multiplying the Hamiltonian
(\ref{1h}) by $r^2$ one redefines the system's energy $E$. For a fixed 
energy in the limit of large $b$, and considering $\theta(0) \approx 0$
it follows from the energy equation that
\begin{equation}
\label{bound}
 |\theta(t)| < \frac{|p_{\phi}| + \sqrt{2 \, \mu \,E}}{e\, b \, r} 
\end{equation}
Therefore for large $b$ and small $E$ we have that $|\theta(t)|$ is small. 
Expanding the Hamiltonian in $\theta$  we obtain
$${\tilde H} = \frac{p_{\theta}^2}{2\, \mu}  +  \frac{(p_{\phi}-
e\,b\,r)^2}{2 \, \mu \,  \cos^2(\theta)}= \frac{p_{\theta}^2}{2\, \mu} 
+ \frac{(p_{\phi}- e\,b\,r)^2}{2 \, \mu} \left( 1 + \theta^2 +
O(|\theta^3|)\right), $$ 
where $\tilde H=r^2\,H$.
We define the ciclotron frequency $\omega_c=\frac{e\,b}{\mu}$. Neglecting third
order terms we quantize ${\tilde H}$ following the standard 
canonical procedure, obtaining the time independent Schrodinger equation
\begin{equation}
\label{ind}
\left\{-\frac{\hbar^2}{2\,\mu}\,\frac{\p^2}{\p\theta^2}+\frac{1}{2\mu}\,
\left(i\hbar\frac{\p}{\p\phi}+e\,b\,r\right)^2 \left(1+\theta^2\right)\right\}
\Psi(\theta,\phi)=E\,\Psi(\theta,\phi).
\end{equation}
We stress that, since we are working in a non-flat, non-contractible manifold, the
quantization
approach is not well defined(\cite{wood}). Therefore, our choice for this procedure
is based on the classical locality argument given above. As it will be
discussed later, our results will show, in the large magnetic field
limit, a good qualitative agreement with the flat Landau problem.
Using the geometry of the sphere and considering the fact that $\phi$ is a cyclic 
variable we impose, 
$$
\Psi(\theta,\phi-\frac{\pi}{2})=\Psi(\theta,\phi+\frac{\pi}{2}),
$$
and write 
$$\Psi(\theta,\phi)=\sum_{n \in \integers} \Theta_m(\theta)\,e^{im\phi}.$$

The equation for $\Theta_m(\theta)$ is given by,
\begin{equation}
\label{tetao}
\frac{d^2}{d\theta^2}\Theta_m(\theta)-\left(
\frac{\mu^2\,\omega_m^2}{\hbar^2} \theta^2 +
\frac{2embr}{\hbar^2}\theta\right)\Theta_m(\theta)
= -\frac{2\mu}{\hbar^2}\tilde E_m \,\Theta_m(\theta) 
\end{equation}
where 
\begin{equation}
\label{omega}
\left\{
\begin{array}{ccc}
\omega_m &=& \sqrt{\omega_c^2+\frac{m^2}{\mu^2}} \\
& & \\
\tilde E_m &=& E - \frac{\hbar^2 \,m^2}{2\mu}. 
\end{array}
\right.
\end{equation}
Writting 

$$
\theta =\rho_m \, y \phantom{/}\mbox{for}\phantom{/} \,  \rho_m =
\sqrt{\frac{\hbar}{\mu\omega_m}},
$$

equation (\ref{tetao}) becomes 
$$
Y_m^{\prime\prime}(y)-\left(y^2+\lambda_m\,y\right)Y_m(y)=-\frac{2\tilde
E_m}{\hbar\omega}Y_m(y)
$$
where
$$
\lambda_m=\frac{2e\,m\,b\,r}{\hbar^2}\cdot 
\left(\frac{\hbar}{\mu \, \omega_m}\right)^{\frac{3}{2}}.
$$
Doing  $x=y+\frac{\lambda}{2}$ we obtain
$$
X_m^{\prime\prime}(x)-x^2\,X_m(x)=-\tilde\epsilon_m X_m(x),
$$
with 
$$\tilde\epsilon_m=\frac{2\tilde E_m}{\hbar\omega}+\frac{\lambda_m^2}{4}. $$ 
This is the 1-dimensional harmonic oscillator equation and therefore one can
determine the energy eigenvalues of our approximated
Hamiltonian obtaining 
\begin{equation}
\label{niveis}
E_{m,l}=\frac{\hbar^2m^2}{2\mu} + 
\hbar\omega\left(l+\frac{1}{2}\right) -\hbar\omega_m\frac{\lambda^2_m}{8}. 
\end{equation}

We observe that for large $b$ the levels $E_{m,l}$ are always positive
since the last term tends to zero. One can also note that we have lost
 the typical degeneracy 
of the Landau problem on  the plane. Due to the compactness of $S^2$ the
eigenvalues of our Hamiltonian  are now labeled by two quantum numbers. 
According to our approximation this expression is valid only on the limit of 
low-energy levels i.e. only for a finite number of eigenstates. Therefore we assume
$\left\{ m,l \right\}  \, < \left\{ m_{\max},l_{\max}\right\}$. The levels
$m_{\max}$
and $l_{\max}$ depend on the magnetic field
value. \par \bigskip
\noindent{\bf Remark:} We observe that in the limit $b \rightarrow \infty$ our
eigenvalues tend to the eigenvalues of the landau problem. In fact
one can compute that $\lim_{b \rightarrow \infty} \lambda_m =0$. Comparing
the magnitudes of the remaining terms for the large field limit we obtain 
the result. This has a simple explanation:
Since the classical solutions of the Landau problem are circles on the sphere,
as the field increases the circles corresponding to the classical solutions contract
to points. Therefore, all classical solutions with small energy concentrate
on a neighborhhod of its initial position and therefore the local flat approximation
of the sphere works better higher is the magnetic field. This is our main
justification for the use of the canonical quantization procedure in this
limit. 
\section{Free Energy Calculation}
\label{mag}

Using the energy eigenstates just computed we calculate the free-energy (\cite{book}), defined by 
$$ F = N \, \nu - \frac{1}{\beta} \, \sum_{m,l} \ln ( 1 + e^{\beta \,\left(\nu - E_{m,l}\right)}) $$
where $\beta = \frac{1}{K \, T}$ and $\nu$ is the 
chemical potential. We redefine the energy in order to consider spin effects
by writing
$$ \tilde E_{m,l} = E_{m,l} \pm \frac{g \hbar \omega_0}{4} ,$$
where $\omega_0= \frac{eb}{m_0}$, with $m_0$ representing the free 
electron mass, 
   $g$ is the effective giromagnetic factor 
and the signs $\pm$ result from different spin directions.
We review briefly how to compute $F$. Introduce the classical partition
function
$$ Z(\beta) = \sum_{m,l} e^{-\beta \, E_{m,l}}$$
and the auxiliary function 
$$ g(E) =( 1 + e^{\beta \,\left(\nu - E \right)}).$$
Let  $\phi(s)$ denote the Laplace transform of $g(E)$. Define
$z(E)$ as the inverse Laplace transfom of $Z(\beta)/\beta^2$ i.e.

$$ \frac{Z(\beta)}{\beta^2} = \int_0^{\infty} z(E) \, e^{-\beta \, E} dE . $$
Therefore
\begin{equation}
\label{zezinho} 
z(E)= \frac{1}{2\, \pi \, i} \, \int_{c-i \, \infty}^{c+i \, \infty} \, e^{E\, s} s^{-2} Z(s) \, ds
\end{equation}
and 
$$ g(E) = \frac{1}{2\, \pi \, i} \, \int_{c-i \, \infty}^{c+i \, \infty} \, \phi(s) \, e^{E\, s} ds. $$
$c$ must be chosen so that all the singularities of the integrals are on the left
of the integration path. With these functions the free-energy can be written as
$$ F = N \, \nu - \beta \sum_{l,m} \, g(E_{l,m}) = N \, \nu - \frac{\beta}{2 \, \pi i}
\int_{c-i \, \infty}^{c+i \, \infty} \, \frac{Z(-s)}{s^2} \, s^2 \phi(s) \, ds $$
since $s^2 \, \phi(s)$ is the Laplace transform of $\frac{\p^2 g}{\p E^2}$ we
obtain finally that
\begin{equation}
\label{free}
 F = N \, \nu - \beta  \int_0^{\infty} z(E) \frac{\p^2 g}{\p
E^2} dE , 
\end{equation}
where
$$ \frac{\p^2 g}{\p E^2} = -\beta \frac{\p f}{\p E} $$ 
for 
$$f(E - \nu) = \frac{1}{1+e^{\beta \, (E - \nu)}} ,$$
the fermi function. Therefore
$$ F = N \, \nu + \int_0^{\infty} z(E) \, \frac{df}{dE} \, dE. $$
We can compute now the unit area partition function for our system:
$$ Z(\beta) = \frac{e \, b}{4 \, \pi \, \hbar^2} \sum_{spin}\,\sum_{l=0}^{l_{\max}}\,
\sum_{m \in \integers}^{m_{\max}}
e^{\pm \frac{\beta \, g \, \hbar \, \omega_0}{4}}\,
e^{-\beta\,(l + \frac{1}{2})\, \hbar \, \omega_m}\, e^{-\beta\,
\left(\frac{\hbar^2 \, m^2}{2 \, \mu}+ \frac{\lambda_m^2\, \hbar \, \omega_m}{2}\right)}
\,.$$ 
Using that 
$$\sum_{l\in\integers} e^{-\beta\,(l + \frac{1}{2})\, \hbar \, \omega_m}\, = 
\frac{1}{\sinh(\frac{\beta\, \hbar \omega_m}{2})},$$
$$ \sum_{spin} e^{\pm \frac{\beta \, g \, \hbar \, \omega_0}{4}} = 2 \, \cosh(\frac{\beta \, g \, \hbar
\, \omega_0}{4}), $$
along with equation (\ref{zezinho}) we obtain 
\begin{equation}
 z(E)=\frac{1}{2\, \pi i}\, \frac{e \, b}{4 \, \pi \, \hbar^2}\,
\sum_{l \in \integers}
\, \int_{c-i \, \infty}^{c+i \, \infty}\frac{e^{-\beta\,\left(\frac{\hbar^2}{2 \, \mu}\,m^2+ 
\frac{\lambda_m^2}{2}\, \hbar \omega_m-E
\right)}\cosh(\beta\frac{\hbar
\omega_0}{4}))}{\beta^2\,\sinh(\frac{\beta\, \hbar
\omega_m}{2})}d\beta\, 
\end{equation}

\begin{figure}
\label{1}
\centerline{\includegraphics[width=60mm,height=60mm]{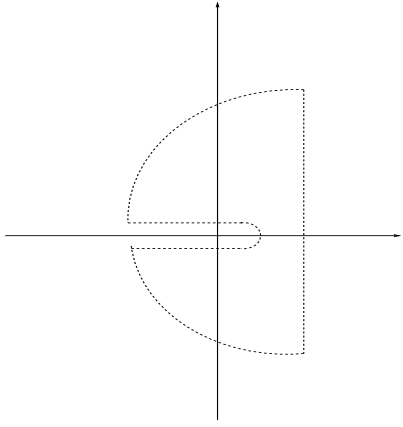}}
\caption{Contour of integration}
\end{figure}

As this integral has poles at 
\begin{equation}
\label{poles}
\frac{\beta\hbar\omega_m}{2}=n\pi i
\end{equation}
for $n\in \integers$, we compute the integral in complex plane, by the convenient path shown in figure
(1)
above. Summing the contribution of each pole in the imaginay axis together with the
small contour around zero, we find
\begin{eqnarray}
z(E)=\frac{eb}{4\pi\hbar^2} \cdot \sum_n\sum_m \left\{ 
\frac{\left(E-\frac{\hbar^2m^2}{2 \, \mu}\,+ 
\frac{\lambda_m^2}{2}\,
\right)^2}{\hbar\omega_m} + 
\left[\frac{1}{2}\left(\frac{g\omega_0}{2\omega_m}\right)^2-\frac{1}{6}\right]\hbar\omega_m
+\dots \right.\nonumber\\
\left.-\frac{\hbar\omega_m}{2} (-1)^n \cdot
\cos\left[\frac{n\pi\omega_0}{2\omega_m}\right]\cdot
\frac{\cos\left\{\frac{2n\pi}{\hbar\omega_m}
\left(E-\frac{\hbar^2m^2}{2 \, \mu}\,+ 
\frac{\lambda_m^2}{2}\,
\right)\right\}}{(n\pi)^2}
 \cdot \right\}\nonumber
\end{eqnarray} 
where $n$ labels the $n$-th pole in the imaginary axis, and the double sum is consistent
with (\ref{poles}).
Recalling that the free-energy is given by (\ref{free}) we get,
\begin{eqnarray}
F = N\nu + \int_0^\infty z(E) \, \frac{\p f(E-\nu)}{\p E} \, dE,
\end{eqnarray} 
and using the Fermi function defined we obtain
\begin{equation}
F=N\nu - \int_0^\infty z(E) \cdot \frac{\beta}{4\cosh^2
\left(\frac{\beta(E-\nu)}{2}\right)} \,dE .
\end{equation}
In order to compute $F$ we proceed in two steps. First we compute the 
following integral:
\begin{equation}
\int_0^\infty \cos\left(\frac{2\,\pi \, E}{\hbar \, \omega_m} - \frac{\alpha_{n,l} }{\hbar \, \omega_m} 
\right)\frac{\beta}{4\cosh^2
\left(\frac{\beta(E-\nu)}{2}\right)} \,dE ,
\end{equation}
where 
\begin{equation}
\label{alpha}
\alpha_{m,n} =\frac{\hbar^2m^2}{2 \, \mu}\,+ 
\frac{\lambda_m^2}{2}.
\end{equation}
Defining $y = \beta (E - \nu)$ we can write this integral as

\begin{equation}
-\frac{1}{4} \,Re \left\{\,e^{i \left(\frac{2\, \pi \,m \nu}{\hbar \,
\omega_m}
-\frac{\alpha_{m,l}}{\hbar \, \omega_m} \right)}\,\,\int_{-\infty}^{+\infty}
\frac {e^{\frac{2\pi n i}{\beta\hbar\omega_m}\,y}}{\cosh^2\left(\frac{y}{2}\right)}\,dy
\right\};
\end{equation}
We have extended the integral lower limit from $-\nu \beta$ to $-\infty$. The error caused by this
replacement is of the order $e^{-\nu\beta}$, which is negligible, considering the low temperature 
limit of our model. This integral can be performed in any algebraic software and the result is given by:

\begin{equation}
-\left(\frac{2\, \pi^2 \beta \,n}{\hbar \, \omega_m}\right)\cdot \frac{\cos
\left(\frac{2\, \pi \,n \nu}{\hbar \, \omega_m}
-\frac{\alpha_{m,l}}{\hbar \, \omega_m} \right)
}{\sinh\left(\frac{2\, \pi^2 \beta \,n}{\hbar \, \omega_m}\right)}.
\end{equation}
For the second step we consider the integral of the terms coming from very 
high temperatures, 
namely, the contributions to the former integral coming from the small contour around
$\beta=0$. Observing that the Fermi function tends to a Heavside step function on the limit of
$\beta\rightarrow 0$, we write $\frac{\p f}{\p E}(E-\nu)=-\delta(E-\nu)$, and the
computations of the high temperature terms are trivial. 
Collecting those results we finally obtain that

\begin{eqnarray}
\label{liv}
F=N\nu&+&\frac{eb}{4\pi\hbar^2} \cdot\sum_m \left\{ 
\frac{\left(\nu-\alpha_{n,m}\right)^2}{\hbar\omega_m}\,+
\left[\frac{1}{2}\left(\frac{g\omega_0}{2\omega_m}\right)^2-\frac{1}{6}\right]
\hbar\omega_m\right\}+ \\
\nonumber
&+&\frac{eb}{4\pi\hbar^2} \cdot\sum_m \sum_n\frac{(-1)^n}{4\beta n}
\frac{\sin\left\{\frac{2n\pi}{\hbar\omega_m}\nu- \frac{\alpha_{n,m}}
{\hbar\omega_m}\right\}\cdot \cos\left[ \frac{n \, \pi \, \omega_0}{2\,
\omega_m}\right]}{sinh\left(\frac{2\pi^2n}{\beta\hbar\omega_m}\right)}\nonumber
\end{eqnarray}
 
\section{Magnetization and the de Haas-Van Alphen effect }
\label{curv}
The magnetization $M$ is found differentiating the free-energy (\ref{liv}) with
respect
to $b$. In the low temperature limit we can write that

$$ F \approx \frac{eb}{4\pi\hbar^2} \cdot\sum_m \sum_n\frac{(-1)^n}{4\beta n}
\frac{\sin\left\{\frac{2n\pi}{\hbar\omega_m}\nu- \frac{\alpha_{n,m}}
{\hbar\omega_m}\right\}\cdot \cos\left[ \frac{n \, \pi \, \omega_0}{2\,
\omega_m}\right]}
{\sinh\left(\frac{2\pi^2n}{\beta\hbar\omega_m}\right)} $$

To compute $M$ we differentiate $F$ and 
observe from (\ref{omega}) that for large $b$,  
$\omega_m \approx b$. We also note that for large $b$ the ratio
$\frac{\omega_0}{\omega_m}$ is of order $0$ in $b$. From
(\ref{alpha}) we obtain that in the large field limit the leading terms
in magnetization per unit of area are given by  
\begin{eqnarray}
\label{deHass1}
M &=& \frac{e}{4\pi\hbar^2} \cdot\sum_m \sum_n\frac{(-1)^n}{4\beta n}
\frac{\sin\left\{\frac{2n\pi}{\hbar\omega_m}\nu- \frac{\alpha_{n,m}}
{\hbar\omega_m}\right\}\cdot \cos\left[ \frac{n \, \pi \, \omega_0}{2\,
\omega_m}\right]}
{\sinh\left(\frac{2\pi^2n}{\beta\hbar\omega_m}\right)} 
\\ \nonumber &\times& \left( 1 + \,\frac{e\,2\, \pi^2 \, n
\omega_c}{\tanh\left(\frac{2\pi^2n}{\beta\hbar\omega_m}\right)\,\mu
w_m^3\, \beta \,  \hbar}\right). \end{eqnarray}
The magnetization is thus expressed in terms of a superposition of periodic functions on
$\frac{1}{\omega_m}$. For the planar case $\omega_m=\omega_c$ and therefore only one frequency 
dominates the oscilatory behaviour of the magnetization. We remark that on the 
$\beta \rightarrow \infty$ limit, which corresponds to the low temperature regime,
 the denominator of (\ref{deHass1}) tends to a constant and we
have a pure superposition of periodic functions. We call this behaviour a {\it de Hass-Van Alphen
type effect}. 

The analogy with the classical Landau problem is better drawn looking at
the ground state of our system: $l=0 \,$,$m=0$.
In this case we have $\omega_m=\omega_c$, $\alpha_{m,n}=0$ and we obtain for (\ref{deHass1})
\begin{eqnarray}
\label{deHass2}
M &=& \frac{e}{4\pi\hbar^2} \cdot \sum_n\frac{(-1)^n}{4\beta n}
\frac{\sin\left\{\frac{2n\pi}{\hbar\omega_c}\nu \right\}\cdot \cos\left[
\frac{n \, \pi \, \omega_0}{2\,
\omega_c}\right]}
{\sinh\left(\frac{2\pi^2n}{\beta\hbar\omega_c}\right)}
\\ \nonumber &\times& \left( 1 + \,\frac{e\,2\, \pi^2 \, n 
}{\tanh\left(\frac{2\pi^2n}{\beta\hbar\omega_c}\right)\,\mu
w_c^2\, \beta \,  \hbar}\right) \end{eqnarray}
Considering that in the low temperature limit $\beta >> 1$ 
the second term in the bracketts of (\ref{deHass2}) tends to $\frac{1}{b}$, we can write in the large
magnetic field limit that 

\begin{equation}
\label{zero}
M = \frac{e}{4\pi\hbar^2} \cdot \sum_n\frac{(-1)^n}{4\beta n}
\frac{\sin\left\{\frac{2n\pi}{\hbar\omega_c}\nu
 \right\}\cdot \cos\left[
\frac{n \, \pi \, \omega_0}{2\,
\omega_c}\right]}
{\sinh\left(\frac{2\pi^2n}{\beta\hbar\omega_c}\right)}, 
\end{equation}
that is up to a phase the exact expression for the Landau planar problem in
two dimensions. 
\section{conclusions}
\label{conc}
From (\ref{zero}) we see that the lowest energy state, as expected, " do not see "
the curvature of the sphere. The complete solution of spherical Landau problem
will be a superposition of different discrete frequencies, each one 
corresponding to a different energy eigenstate. We remark that, as explained
before, (\ref{deHass1}) holds only in the limit of low energy states. Also we
must stress that this result is valid in the range where the flat de Haas-van
Alphen is also observed, i.e. in the low temperature-high magnetic field regime.

A more general treatment of the spherical Landau  problem still remains to be done,
in the sense that one should consider the other limiting cases. Due to the non-local
aspects of those limits, more sophisticated techniques are to be employed. Whatever
such approach might be, we expect that our present results arise as an assymptotic,
low temperature-high magnetic field, behaviour.

\end{document}